\documentclass[oldversion]{aa}%[referee]

\usepackage{natbib}
\usepackage{graphicx}
\usepackage[mathcal]{eucal}
\usepackage{amssymb} 
\usepackage{revsymb}	
\usepackage{amsmath}
\usepackage[usenames]{color}	
\usepackage{epstopdf}	
\newcommand{\ds}{\displaystyle}

\long\def\jumpover#1{{}}

\newcommand{\eq}[1] {Eq.~(\ref{#1})}

\newcommand{\fig}[3]{
      \begin{figure}[ht]
        \begin{center}
        \resizebox{\hsize}{!}{\includegraphics  {#1}}
        \end{center}    
        \caption{#2}
        \label{#3}
        \end{figure} }
\newcommand{\eqn} [1] {
\begin{equation}#1
\end{equation}}

\usepackage[applemac]{inputenc}

\def\acenA{$\alpha$~Cen~A}

\newlength{\lenA} % 
\begin{document}

\title{Modelling the excitation of acoustic modes in {\acenA}}

\author{
Samadi R. \inst{1}    
\and K. Belkacem\inst{1}  
\and M.J. Goupil\inst{1}    
\and M.-A. Dupret\inst{1}  
\and F. Kupka\inst{2}
}
\institute{
Observatoire de Paris, LESIA, CNRS UMR 8109, 92195 Meudon, France \and 
Max-Planck-Institute for Astrophysics, Karl-Schwarzschild Str. 1, 85748 Garching}

\offprints{R. Samadi}
\mail{Reza.Samadi@obspm.fr}

\date{\today} % Received / Accepted}

\titlerunning{.}

%%%%%%%%%%%%%%%%%%%%%%%%%%%%%%%%%%%%%%%%%%
\abstract{
We infer from different seismic observations the energy supplied
  per unit of time by turbulent convection to the
acoustic modes of {\acenA} (HD 128620), a star which is
similar but not identical to the Sun. 
The inferred rates of energy supplied to the modes ({\it i.e.} mode excitation rates) are found to be
significantly larger than in the Sun.
They are compared with those computed with an
 excitation model that includes two sources of driving, the Reynolds stress contribution and the 
advection of entropy fluctuations.
The model also  uses a  closure model, the Closure Model with Plumes (CMP hereafter), that takes the asymmetry
between the up- and down-flows ({\it i.e.} the granules and plumes, respectively) into account.
Different prescriptions for the eddy-time correlation function are
also confronted to observational data.
%The comparison between theory and observations
% allow us to derive 
%physical constraints on turbulent convection for {\acenA}, 
%can be derived from the comparison between theory and the observations for on
%other star than the Sun.
Calculations based on a Gaussian eddy-time correlation
underestimate excitation rates compared with the values derived from observations for
{\acenA}. On the other hand, calculations based on a Lorentzian eddy-time
correlation  lie within the observational error bars. This confirms
results obtained in the solar case.
With respect to the helioseismic data, those obtained for {\acenA}
constitute  an additional support for our model of excitation. 
We show that mode masses must be computed taking  turbulent pressure into account.
Finally, we emphasize the need for more accurate seismic measurements in order to
discriminate, in the case of  {\acenA}, between the CMP closure model and the quasi-Normal
Approximation as well as to confirm or not the need to include the
excitation by the entropy fluctuations. 
}

\maketitle

%%%%%%%%%%%%%%%%%%%%%%%%%%%%%%%%%%%%%%%%%%
\section{Introduction}

{\acenA} is the most promising star after the Sun for constraining
the modelling of $p$-mode excitation by turbulent convection.
Indeed, due to its proximity and its binarity, the fundamental parameters of {\acenA}
(effective temperature, 
luminosity, metallicity, gravity, radius) are quite well known.
For this reason this star and its companion ($\alpha$ Cen B)
have been extensively  studied (see for instance the most recent modelling by
 \citet{Miglio05} and the references therein).
As pointed out recently by \citet{Samadi07b}, the detection of $p$-modes and the measurement of their 
amplitudes as well as their mode line-widths ({\it i.e.} lifetime)
from {\acenA} enable to derive the rates at which energy is
supplied to the acoustic modes for this star. These observational constraints   can then  be used to check  
 models of $p$-mode excitation by turbulent convection.

Such comparisons have been first undertaken in the case of the Sun by different
authors \citep[see  the recent review by][]{Houdek06}.
%{GMK94,Samadi00II,Samadi02II,Chaplin05,Samadi05b,Kevin06c,Samadi07a}. 
They enable to test different models of stochastic excitation of
acoustic modes as well as different models of turbulent convection 
\citep[see eg.][]{Samadi05b}.
Among those theoretical prescriptions, we consider that of \cite{Samadi00I} 
with the improvements proposed by \cite{Samadi02II} and \cite{Kevin06b}.
It was shown by \cite{Samadi02II} that the way the
eddy time \emph{correlation}  is modelled plays an important
role on the efficiency of excitation. 
Indeed, calculations of the mode excitation rates, $\mathcal P$, 
that use a Lorentzian eddy-time correlation function
better reproduce helioseismic data than those 
using a Gaussian one. 
%Investigation of the closure models needed in the 
%theoretical description of excitation rates has 
% demonstrated that asymmetries in the convective upward and downward motions 
%is a key feature in the excitation mechanism.
%As verified by \cite{Kevin06a}, the presence of two flows 
%introduces an additional contribution when averaging            
%the fluctuating quantities, since averages of fluctuating 
%quantities over each individual flows differ from averages over the total flow. 
In addition, \cite{Kevin06b}, in the case of the Sun, showed that excitation rates 
computed using an adapted closure model that takes the presence of plumes into 
account reproduce much better the solar
observations than the calculations based on the classical Quasi-Normal
Approximation \citep{Million41}.

An alternative theoretical model of the excitation of acoustic modes by
turbulent convection  proposed by \citet{Chaplin05} differs from that
by \citet{Samadi00I} in several ways:
it does not take the driving by the advection of the entropy
fluctuations by the velocity field into account. They
only use the classical Quasi-Normal Approximation.
More importantly, these authors claim that a Gaussian eddy-time correlation
function reproduces better than a Lorentzian one the frequency dependence of mode excitation
rates inferred from helioseismic data.
However, they are led to introduce in their model a factor by which
they multiply their formulation in order to reproduce the maximum of
the  solar mode excitation rates.

{\acenA}  provides a second opportunity to test various
  assumptions in the 
   modelling of the p-mode excitation: 
the  amplitudes of the  acoustic modes detected in {\acenA} were derived by 
\cite{Butler04} using spectrometric data. 
From those data,  an estimate of the averaged mode line-widths has been
first proposed by \cite{Bedding04} and more recently updated in \cite{Kjeldsen05}.
Using a different method and data from the WIRE satellite,
\cite{Fletcher06} proposed a new estimate of the
averaged mode line-widths that differ significantly from 
 the one derived by \cite{Kjeldsen05}.
Indeed, the two data sets place the mode life time between 2.2 days \citep{Kjeldsen05}
  and 3.9 days \citep{Fletcher06}. For comparison, the averaged mode life time derived for the Sun by
\cite{Bedding04} in a similar way as for {\acenA} by
\cite{Kjeldsen05} is about two days. 
%Hence, p-modes in {\acenA} have
%larger  mode life times than in the Sun.

\citet{Samadi07b} have inferred from those sets of seismic
  constraints  the p-mode
excitation rates $\mathcal P$. They have found that they are
significantly  larger
than those associated with the solar p-modes. 
Furthermore, $\mathcal P$ peaks  in the frequency domain $~\sim $
2.2  --  2.6~mHz  while it peaks at the frequency 
$\nu_{\rm max} \sim $~3.8 mHz in the case of the Sun.

Although the spectroscopic characteristics ($T_{\rm
  eff}=5810$~K, $\log~g =$~4.305) of {\acenA}  are close to that of the Sun ($T_{\rm
  eff}=5780$~K, $\log~g =$~4.438), the seismic signatures are quite different.
Consequently, finding agreement
between predicted and observed excitation rates would be a non-trivial
result, providing additional support for the theory.

A preliminary comparison with theoretical calculations obtained in the
manner of \citet{Kevin06a} was carried out  by \citet{Samadi07b}. 
Discrepancies between the excitation rates inferred from the 
observations and the theoretical calculations were found. 
%By going a step further in the modelling, as detailed in Sect.~2.3, 
We update here this study by proceeding in a similar way as
\citet{Rosenthal99}. Indeed, these authors have built a solar 1D model
where the surface layers are taken directly from a  fully compressible
3D hydrodynamical numerical  model. We will refer here to such a 1D model as
a ``patched'' model. 
\citet{Rosenthal99} have obtained a much better agreement between
observed and theoretical eigenfrequencies of the Sun computed for such
a ``patched'' 1D model than those obtained for a ``standard'' 1D model based on the
standard mixing-length theory with no turbulent pressure included. 
Following \citet{Rosenthal99}, we build here such a ``patched'' model to
derive  adiabatic mode radial eigen-displacements ($\xi_r$) and mode inertia
($I$). 
We use them to compute the mode excitation rates, which we compare with excitation
rates computed using  $\xi_r$ and $I$ obtained with a ``standard'' 1D model. 

%by using in addition mode eigenfunctions
%calculated in a more consistent way than previously done.

The paper is organized as follows: in Sect.~\ref{modelling} we
describe our procedure to compute the  mode excitation  rates  for the specific
 case of {\acenA}.  
We then describe in Sect.~\ref{constraints} the way the mode excitation rates are
inferred from seismic observations of {\acenA}.
In Sect.~\ref{comparison}, we compare  theoretical calculations of
$\mathcal P$  with those  inferred from the seismic data obtained for {\acenA}.
We compare and explain in Sect~\ref{differences} the differences
between {\acenA} and the Sun.   
Finally, Sect.~\ref{discussion} and Sect.~\ref{conclusion} are devoted to the discussion and
conclusions, respectively.

%%%%%%%%%%%%%%%%%%%%%%%%%%%%%%%%%%%%%%%%%%
\section{Modelling the excitation of p-modes}
\label{modelling}

%---------------------
\subsection{General formulation}

The theoretical model of stochastic excitation  
is basically that of \citet[][see also \citet{Samadi05c}]{Samadi00I} with the improvements of 
\cite{Kevin06a,Kevin06b}, we thus recall here only some key features. 
The model takes two driving sources into account. 
The first one is related to the Reynolds stress tensor and as such
represents a mechanical source of excitation.  The second one is
caused by the advection of the turbulent fluctuations of entropy by
the turbulent motions (the so-called ``entropy source term'') and thus
represents a thermal source of excitation \citep{GMK94,Stein01II}. 
The power fed into each radial mode, ${\mathcal P}$, is given by:
\begin{equation}
\label{power}
{\mathcal P} = \frac{1}{8 ~ \ds{I} } \left ( C_R^2 + C_S^2  \right )
\;,
\end{equation}
where  $C_R^2$ and $C_S^2$ are the turbulent 
Reynolds stress and entropy contributions, respectively and 
%$\ds{I} = \int_0^M d m \, \| \xi_r \| ^2 $
\eqn{I = \int_0^M d m \, \vert \xi_r \vert^2  
\label{inertia}
}
is the mode inertia, $\xi_r$ is the adiabatic radial mode displacement and $M$ is the mass of
the star.
The expressions for $C_R^2$ and $C_S^2$ are given  for a radial mode with
frequency $\omega_0$ by
\begin{eqnarray}
\label{C2R_ref}
C_R^2 & =  & { 64 \pi^{3} \over 15 } \int d{\rm m}  \,  \rho_0 \,  f_r \, \left(1 + \frac{1}{3} {\cal S}_w^2 \right) \, S_R(\omega_0) \; ,\\
\label{C_S_ref}
C_S^2 &=& \frac{16 \pi^3}{3 \, \omega_0^2}  \int d{\rm m}  \, {
  \alpha_s^2  \over \rho_0 } \,  g_r  \, \, S_S(\omega_0)
\end{eqnarray}
where we have defined
\begin{eqnarray}
\label{SR}
S_R(\omega_0) &=& \,\int  \frac {dk} {k^2 }~E^2(k) ~\int d\omega ~\chi_k( \omega + \omega_0) ~\chi_k( \omega ) \\
S_S(\omega_0) & = & \int \frac{dk}{k^2}\,E(k) \, E_s(k) \, \int
d\omega \, \chi_k(\omega+\omega_0)\,  \chi_k(\omega)
\label{SS}
\end{eqnarray}
where $f_r \equiv (d\xi_r / d r)^2$ and $g_r$ is a function that
involves the first and the second derivatives of $\xi_r$.  $E(k)$ is
the spatial turbulent kinetic energy spectrum, $E_s(k)$ the spectrum
associated with the entropy fluctuations, $\chi_k$ 
the time correlation function of the eddies, $\alpha_s= (\partial P /
\partial s)_\rho$ where $s$ is the entropy, $P$ the gas pressure, $\rho$
the density, $\rho_0$ the equilibrium density profile and
$\omega_0$ the eigenfrequency. 

Finally,  ${\cal S}_w \equiv <w^3>/(<w^2>)^{3/2}$ is the skewness and  $w$
the vertical component of the velocity \citep[see][for details]{Kevin06a,Kevin06b}.  
Indeed, the expression of \eq{C2R_ref} depends on the closure model used to express 
the fourth-order moments involved in the theory in terms of the second
order ones.
The most commonly used closure model at the level of fourth-order moments 
is the quasi-normal approximation (QNA). Such an assumption leads to a 
vanishing skewness ${\cal S}_w$. However, 
% which is valid for a Gaussian probability 
%distribution function (PDF) and was first introduced by \cite{Million41}. 
%In that assumption, the skewness ${\cal S}_w$ vanishes since the PDF is symmetric. 
%But as shown  by  \cite{Kraichnan57} in the context of turbulent flows, 
%the cumulant (which is the deviation from the QNA) can be large and 
%therefore not negligible. 
in the solar case, the deviation from the QNA is due to the 
presence of turbulent plumes. 
%fact that 
%in the uppermost part of the solar convection zone, 
%radiative cooling is responsible for the formation of turbulent plumes, 
%hence the medium is characterized with up- and down- drafts.
By taking both the effect of the skewness introduced by the
presence of two flows and the effect of turbulence onto each flow into
account, \cite{Kevin06a}
thus proposed a new closure model that takes into account the presence of plumes, 
leading to a non-vanishing skewness, ${\cal S}_w$, in \eq{C2R_ref}.
In the present work  ${\cal S}_w$ is then obtained directly from the
3D numerical model.

Calculation of the mode excitation rates is performed essentially in
the manner of \citet{Kevin06b} as
detailed in \citet{Samadi07b} in the specific case of {\acenA}:
 all required quantities, except the mode eigenfunctions $\xi_r$ and mode
inertia $I$, are directly obtained from a 3D simulation of the
outer layers of {\acenA} which characteristics are described in
Sect.~\ref{3Dmodel} below.

The mode displacement $\xi_r$ and mode inertia $I$ must  be computed from a global 1D
equilibrium model. We choose to study two such equilibrium models
which are described in Sect.~\ref{1Dmodels}
Finally, eigenfrequencies and eigenfunctions are computed  using  the adiabatic pulsation code ADIPLS
\citep{JCD91b}.  

%---------------------
\subsection{The 3D hydrodynamical model of the outer layers of
  {\acenA} }
\label{3Dmodel}

We consider in this work the 3D hydrodynamical model of the outer
layers of {\acenA} computed by \citet{Samadi07b}  using
\citet{Stein98}'s code. The assumed micro-physics ({\it e.g.} the
equation of state and the opacity table) are detailed in
\citet{Samadi07a}. The hydrogen, helium and metal abundances are  solar
and the chemical mixture of the heavy elements is set according to \citet{GN93}'s mixture.

The 3D model associated with {\acenA}  has a horizontal size of 8.17
Mm x 8.17 Mm and a vertical size of 4.31 Mm. The grid is $125
\times 125 \times  82$.
As pointed out by \citet{Samadi07a}, this spatial resolution is sufficient for
the calculation of the p-mode excitation rates.
The simulation duration  is 323 minutes while the acoustic depth of
the simulation is 410~seconds and the characteristic eddy turn-over
time is $\sim $ 20 minutes (see Sect.~\ref{differences}). 
The duration of the simulation
then represents $\sim$~47 sound travels  across the simulated domain
and about 15 eddy turn-over times.

%, which duration covers $\sim$
%300 sound travels accross a pressure scale height ($H_p$).

The effective temperature $T_{\rm eff}$ is adjusted to 5809 K $\pm$ 15 in
good agreement with the value $T_{\rm eff}=5810$ K $\pm$ 50 adopted
by  \citet{Miglio05}. The gravity is set to $\log~g~$=~4.305 to match
exactly the value ($\log~g~$=~4.305 $\pm$ 0.005) inferred from the
precise measurements of the mass and the radius of the star
\citep[see the related references in][]{Miglio05}.

%---------------------
\subsection{1D models}

\label{1Dmodels}

%We build two different equilibrium models. We describe below those 1D models.

\subsubsection{Standard model}
\label{standard}
The first  1D equilibrium model has the effective temperature and the 
  gravity of {\acenA} and is built by imposing that for the temperature at the bottom of the
3D simulation box, the 1D model has the same pressure and density as
the  3D simulation (see Fig.~\ref{figC}).  Hence, the 3D simulation
  is used to constrain this 1D 
    equilibrium model such that its interior structure is
    compatible with the second 1D model described later on,
    in Sect.~\ref{patched} (see also Fig.~\ref{figC}).
Convection in the 1D model is 
described according  to  \citet{Bohm58}'s  mixing-length local theory of convection (MLT)
and   turbulent pressure is ignored. Microscopic diffusion of helium and
heavy elements are treated according to the simplified formalism of
\citet{Michaud93}.  We assume a solar abundance in order to be
  consistent with the 3D model.

 The  mixing-length parameter,
  $\alpha$, the age, the mass ($M$), the initial helium abundance ($Y_0$) and the initial
  $(Z/X)_0$ ratio where $X$ and $Z$ are  the hydrogen and metal mass
 fractions, respectively, are fitted such that the model 
  simultaneously reproduces
   the effective temperature of the star, its gravity, the solar
  composition and the   temperature-pressure relation at the bottom of the 3D simulation
  box.  The outer layers of this model, which matter here, then
    have the stratification given by a standard MLT model. This model
    will be referred to as {\bf standard} hereafter. 

 The matching results in $\alpha= 1.694$. For comparison, the
same matching performed for a solar 3D simulation results in
$\alpha= 1.899$.
The mass of the standard model is $M=1.012~ M_\odot$ and the radius $R=1.1722~ R_\odot$.
They are sligthly smaller than that expected for this star, namely $M=1.105
~ \pm 0.007 ~M_\odot$ and
$R=1.224 ~ \pm 0.003 ~R_\odot$ \citep[see][]{Miglio05}.
This is because we have assumed a solar abundance for consistency with
the 3D model.
A global 1D model with the  iron to hydrogen abundance ([Fe/H]) of {\acenA} (namely
[Fe/H]=0.2), would have the expected mass and radius of the star.

Slightly different $R$, $M$ and [Fe/H] values might have some influence on the mode
excitation rates (${\mathcal P}$).
In order to measure the effect of having an $R$, $M$ and [Fe/H] different
than required for {\acenA}, we have computed two global models.
The first model has an abundance [Fe/H]=0.2 and the second  has a solar abundance.
Both models have the effective temperature and gravity of {\acenA}.
In contrast  with the ``standard''  model described above, we do not match
these models with the 3D model. 
The model with [Fe/H]=0.2 has almost the radius and the mass expected for
{\acenA} while the second has almost the same $R$ and $M$ as the
standard model investigated here.
We find that  ${\mathcal P}$  changes between the two
models by less than $\sim 5$\%; this is much smaller than the uncertainties
associated with the observations.

\subsubsection{``Patched'' model}
\label{patched}
\begin{figure}[ht]
        \begin{center}
        \resizebox{\hsize}{!}{\includegraphics  {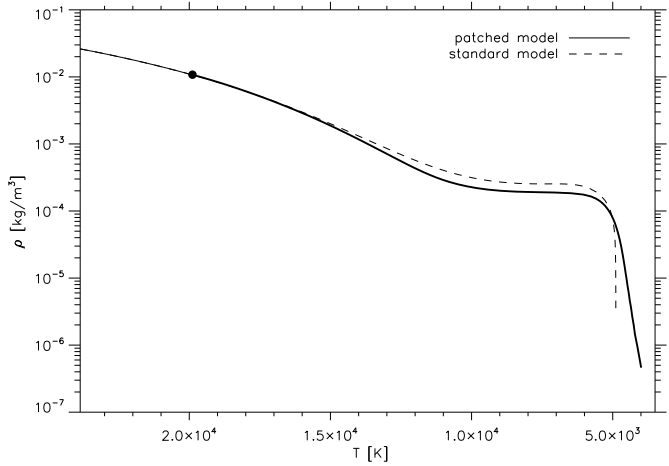}}
        \end{center}    
\caption{ Density as a function of the
  temperature. The solid line corresponds to the ``patched'' model
 and the dashed line to the ``standard'' model. The thick solid line
 is the part of the patched model obtained  from the 3D
 simulation. The filled circle shows the position of the bottom of the
 3D simulation box.
}
\label{figC}
\end{figure}

To take into account a more realistic description of the superadiabatic outer layers, 
we have built, following  \citet[see also \citet{Samadi07a}]{Trampedach97}, a 1D global
model in which the outer layers are replaced by the averaged 3D simulation
(see Fig.~\ref{figC}). 

 ``Standard''  and 3D models  share the same micro-physics but
mainly differ in the way convective motions and radiative transfer are
treated. In the 3D model convective motions are treated by solving the
Navier-Sokes equation while  in the standard model convective motions
are modelled according to the mixing-length model of convection and no
turbulent pressure is included in the hydrostatic equation.
Concerning the radiative transfer, in the standard model radiative
transfer is  gray and assumes the Eddington approximation. In the 3D
model radiative transfer is explicitly solved in LTE for four opacity
bins \citep[see details in][]{Stein98}.

The interior of the ``patched'' model is the same as in the standard model and
does not include the turbulent pressure. At the bottom of the
3D simulation box, turbulent pressure is already negligible
($\sim$~0.6~\% of the total pressure). Then,
neglecting it in the interior has negligible effects on the
properties of the eigenfunctions considered here.
This global model will be referred to as a  {\bf patched}  model. 
 Note that this patched model has the same total mass and a radius very close to that of
  the standard model, namely $R=1.1726~R_\odot$.

The  stratifications in density of the patched and standard models are
compared in Fig.~\ref{figC}.
At the top of the convective region, we see that the density is lower
in the patched model compared to the standard model.
This is because the patched model includes turbulent  pressure that
provides additional support  against gravity: 
accordingly, at a given total pressure ($P_{\rm tot}$), the
patched model has a lower gas pressure. 
Now, since $T(P_{tot})$ is the same with and without turbulent pressure,
at a given temperautre the patched model has a lower density
compared to the standard model.

Because  the treatment of photospheric radiative transfer is not (and
in practice cannot be)   identical between 
a 3D calculation and a 1D  model in the atmosphere, small differences
in  the stratifications of the very outer layers
exist between the two models  as seen in Fig.~\ref{figC}. 
In any case these differences do not play any significant role on the
quantities which matter here, such as inertia. 

Note that some explanations about the differences seen between the outer
layers of 3D models and 1D models 
have been proposed by {\it e.g.} \citet{Nordlund99b} and \citet{Rosenthal99}.

%On the other hand, the density  in the upper part of the atmosphere is -~at a given
%temperature~- higher in the patched atmosphere.
%{\bf This difference  is related to the fact that the atmosphere is
%  treated differently between the 1D standard model and the 3D model
%  (see above).
% }

% At the top of the
%convective region, the density is lower as a result of the 
%presence of turbulent pressure.  
%height $H_p \equiv P_0/(\rho_0 g_0)$, where 
%$g_0$ is the mean gravity which is the same in the patched and in the
%standard model.  Hence, at the same pressure, the pressure scale $H_p$
% is larger in the patched model than in the model with no
%turbulent pressure. 
%This results for the patched
%model in an ``elevation'' of the surface \citep[see eg.][]{Rosenthal99}. Indeed, because of the larger
% $H_p$, the atmosphere extends higher in the patched
%model. Hence, height in the atmosphere, density remains larger in the
%patched model compared to the standard model. 

%%%%%%%%%%%%%%%%%%%%%%%%%%%%%%%%%%%%%%%%%%
\section{Inferring the excitation rates from seismic constraints} 

\label{constraints}

Mode excitation rates are derived from seismic observations
according to the relation:
\eqn{
\mathcal P(\nu) =  2 \pi \, \mathcal M \, \Gamma \, (v/S_0 )^2 
\label{Pobs}
}
where $\ds{{\mathcal M}= I /
  \xi_r^2(r_h)}$ is the mode mass evaluated at the layer 
 $r_h \equiv R + h$ in the atmosphere where the mode is measured in
radial velocity,
 $R$  the radius at the photosphere ({\it i.e.} at $T=T_{\rm
  eff}$), $h$ the height above the photosphere, $\Gamma$ the
mode full width at half maximum (in $\nu$), 
$v(r_h,\nu)$ is the rms \emph{apparent} velocity amplitude of the mode at the
layer  $r_h$, $\nu = \omega_0 / 2 \pi$ the mode frequency  and  $S_0$
the visibility factor of the $\ell$=0 mode.

\citet{Kjeldsen05} have derived the apparent amplitude velocity spectrum,
$v(\nu)$ of the modes detected in {\acenA}.  
However, their spectrum  corresponds to amplitudes
normalized to the mean of $\ell$=0 and 1 modes rather than to $\ell$=0. 
Furthermore, they do not take into account the mode visibilities.
Recently, \citet{Kjeldsen08} have derived the 
(apparent) amplitudes of the modes, normalised to the $\ell$=0 modes
and taking into account both the mode
visibilities and limb-darkening effects.
Finally, in order to derive the {\it intrinsic} mode
amplitudes, we divide $v(\nu)$ by $S_0= 0.712$, the visibility factor of
the $\ell$=0 modes observed in velocity \citep[][]{Kjeldsen08}. 

For the mode line-width, $\Gamma$, we use  the
averaged values  provided  by  \citet{Kjeldsen05} and
\citet{Fletcher06}.

Concerning mode masses, $\mathcal M$, as discussed in Sect.~\ref{discussion}, it is not trivial to determine the height
  $h$ where the Doppler velocities are predominantly measured.
As we do not precisely know  the height representative of the
observations, we  will next evaluate the mode masses --~by default~--
at the optical depth $\tau_{\rm~500 nm} \simeq 0.013$, which corresponds to the
 depth where the potassium (K) spectral line is formed (but see
Sect.~\ref{discussion} for the discussion).
This optical depth corresponds to $h$=470~km.

 Neither the standard nor the patched models do  have exactly the
  radius and the mass expected for {\acenA} (see Sect.~\ref{standard}).
However, this inconsistency has only negligible effect on the mode mass
${\mathcal M} =I/\xi_r^2$. Indeed, since the eigenmode displacement, $\xi_r$, is directly
proportional to  $R$, the mode inertia $I$ scales as $R^2$ (see
Eq.~(\ref{inertia})). Accordingly, the   ratio $I/\xi_r^2$ is almost
insensitive to a small change in $R$. 
Furthermore, we have checked that ${\mathcal M}$ is also insensitive to
a small change in $M$.

%%%%%%%%%%%%%%%%%%%%%%%%%%%%%%%%%%%%%%%%%%%%%%%%%%%%%
\section{Comparison between observations and modelling} 
\label{comparison}

We first compare theoretical calculations of $\mathcal P$ performed using
eigenfunctions computed with the {\it patched} equilibrium model 
with those computed using the {\it standard} equilibrium model (see Sect.~2.2).  
However, eigenmodes computed with those two models have not the same
inertia and hence not the same mode masses $\mathcal M$.
Thus, we rather compare the ratios ${\mathcal P}/{\mathcal M}$.
As shown in Fig.~\ref{figA}, theoretical calculations that use the
{\it patched} model lie well inside the observed domain of the seismic constraints. On
the other hand, using the
{\it standard} model   leads to  underestimated theoretical values compared to the  two sets of 
seismic constraints. 

When comparing the  integrands of the {\it product}  ${\cal P} {\cal M}$ --~
excitation power times mode mass ~--  between 
 the patched and standard equilibrium models, we find that they are  quite
similar. On the other hand, the mode masses ${\cal M}$ 
are quite different for the two equilibrium models in the domain 1-3 mHz where the modes
are mostly excited. This is due to the turbulent pressure which is present in the patched model and
ignored in the standard model.  At a given radius in the super-adiabatic region, 
the patched model has a lower gas pressure and density (see Fig.~\ref{figC}). 
As a consequence,
 inertia of the modes, which are confined within the super-adiabatic region, 
 where the turbulent pressure has its maximum, are smaller  for the patched model 
  than for  the standard model; accordingly,  the {\it ratios} ${\cal P}/{\cal M}$,
which are inversely proportional to  the squared mode mass ${\cal
  M}^2$, are  about two times larger  for the patched
model.

In Fig.~\ref{figB}, we compare two sets of calculations for a patched stellar model 
that assumes two different prescriptions for the eddy-time correlation function
($\chi_k$) and two different closure models, namely the QNA and the CMP in the excitation model.
The theoretical calculations based on a Lorentzian $\chi_k$ and the
CMP closure model lie inside the range allowed by the two sets of seismic constraints.
The differences between calculations based on the CMP
 and on the QNA are found smaller than the differences between
 the two data sets. 
On the other hand, calculations based on a Gaussian $\chi_k$ yields 
significantly underestimated values compared to the seismic
constraints.

Note that \citet{Samadi07b} found a discrepancy between theoretical calculations and observations.
Part of this discrepancy  was due to the fact that the horizontal size of the
simulation box was set to that of the solar simulation
used in \citet{Kevin06b}. % not at that time correctly set.
 Indeed, the kinetic energy
spectrum $E$ involved in the expression 
for $S_R$ and $S_S$ in Eqs.~ (\ref{SR}) and (\ref{SS}) must be 
normalised with respect to the
horizontal size of the simulation box, as is done here. 
 Furthermore, the mode inertia considered by \citet{Samadi07b} were
  computed  for  a standard MLT model ({\it i.e.} no
  turbulent pressure included) instead of using a
  patched model as is done here. As  shown in
  Fig.~\ref{figA},   this results in an under-estimation of the ratio
  ${\mathcal P / \mathcal M}$ by a factor of about two (see Fig.~\ref{figA}).

\fig{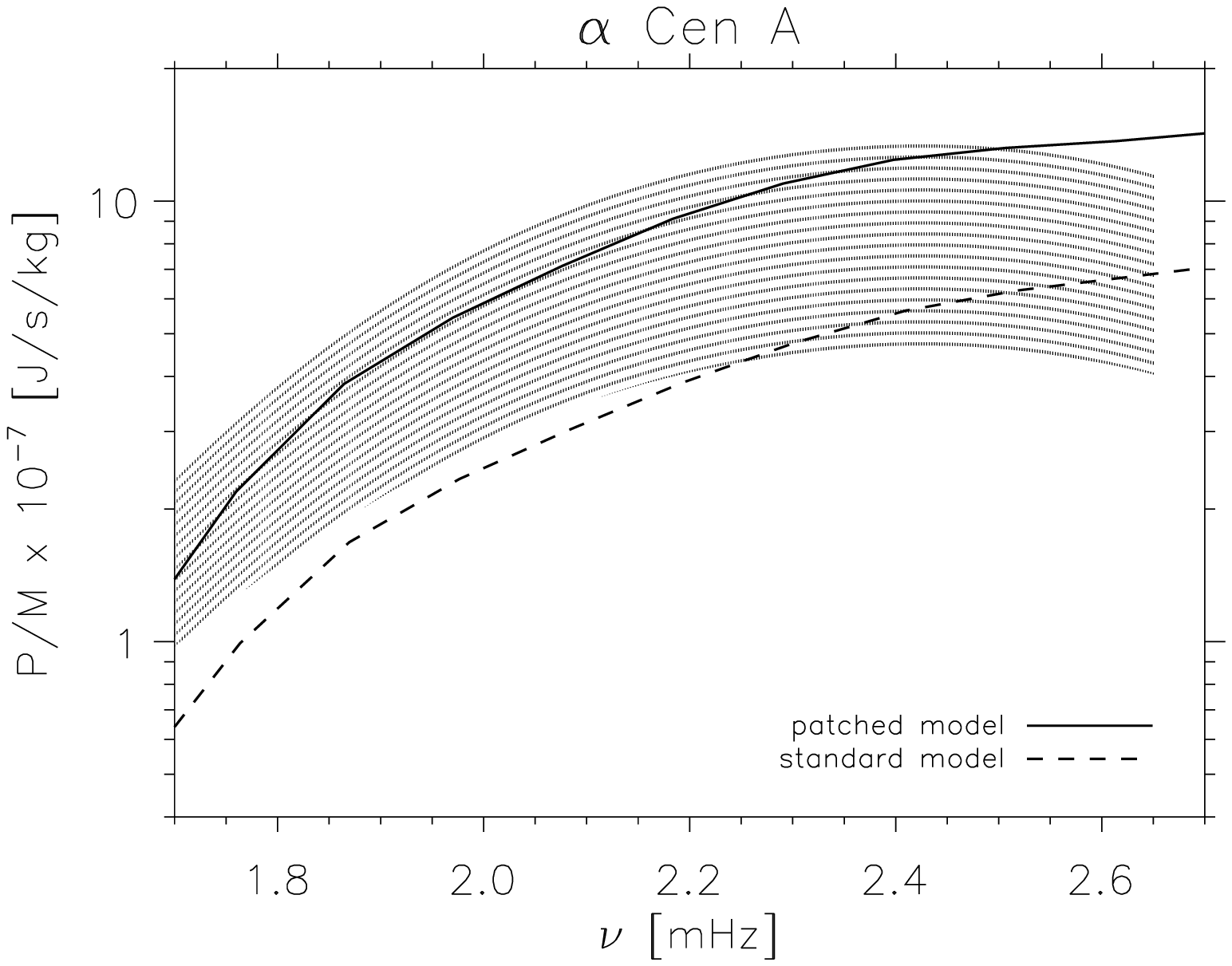}{Ratio of the rates ${\mathcal P}$ at which energy is injected into
   $p$-modes   to mode masses (${\mathcal
    M}$) for {\acenA}. 
The dashed area represents the observed domain for ${\mathcal
  P}/{\mathcal M} = 2 \pi\, \Gamma \,  (v/S_0)^2 $  as a function of $\nu$. This domain is defined 
by merging the  uncertainties associated with  two independently derived values of
$\Gamma$ and with  the mode amplitudes $v$ (Eq.~\ref{Pobs}).
The solid (resp. dashed) line corresponds to computed excitation rates
with the eigenmodes obtained using
    the ``patched'' (resp. ``standard'') 1D global model.
  All calculations here
    use the CMP and the  Lorentzian function (LF) for the  eddy
    time-correlation  function $\chi_k$ in Eqs.~ \ref{SR} and \ref{SS}.
}{figA}

\fig{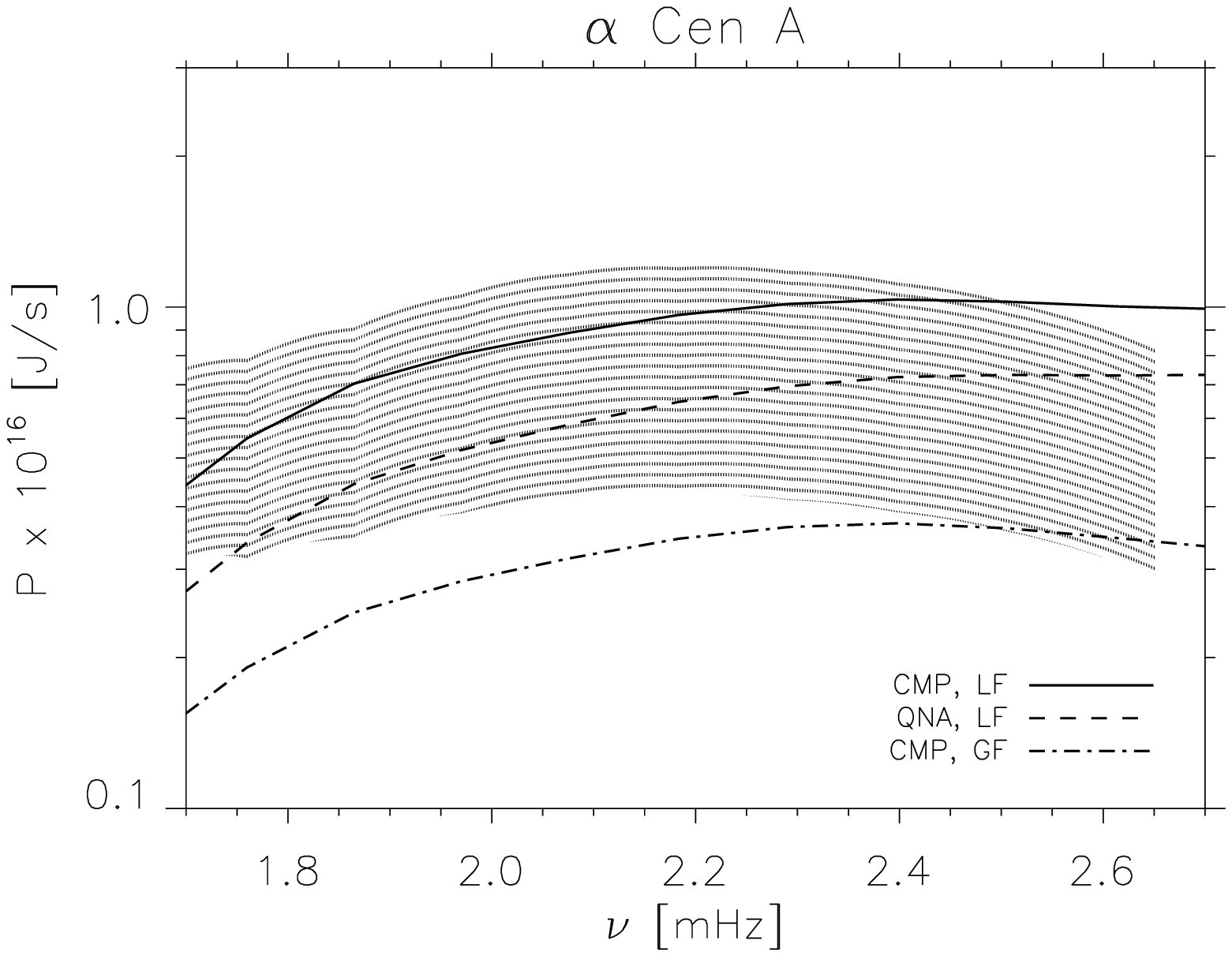}{Rates $\mathcal P$ at which energy is injected into
  the $p$-modes of {\acenA}. 
The dashed area has the  
  same meaning as in Figure~\ref{figA}.  
 The  lines correspond to different theoretical
    calculations (all using a patched model): 
    the solid line uses the Lorentzian function (LF) and the CMP,
    the dashed line uses the LF and the QNA closure model, the
    dot-dashed line uses the Gaussian function (GF) and the CMP. }{figB}
%\fig{plot/acenA_cmp_pow.1.eps}{}{figC}

%\fig{plot/acenA_cmp_pow.2.eps}{}{figC}

%%%%%%%%%%%%%%%%%%%%%%%%%%%%%%%%%%%%%%%%%%
 
\section{Differences between {\acenA} and the Sun}

\label{differences}

%---------------------------
\subsection{Excitation rates}

Fig.\,\ref{figH} compares the excitation rates, ${\mathcal P}$,
inferred for {\acenA} with those inferred
from  helioseismic measurements obtained for the Sun.
For  {\acenA}, excitation rates are obtained from the seismic
measurements as explained in Sect.~\ref{constraints}.

For the Sun we consider the helioseismic data studied by
\citet{Baudin05}. We use here solar mode masses obtained with a patched model
computed as for {\acenA} in Sect.~\ref{modelling}.
Mode masses are evaluated for the optical
depth $\tau \simeq 5~10^{-4}$ since SOHO/GOLF observations are based on the Na
D1 and D2 spectral lines \citep[see][]{Houdek06}.

We find $\mathcal P_{\rm max,\odot} \simeq 
3.5 \, \pm \, 0.4 \,\times \,  10^{15}$ [J/s].
The excitation rates inferred for  {\acenA} with mode masses $\mathcal
M$ evaluated at 
the optical depth associated with the potassium line ($\tau \simeq$
0.013) 
give $\mathcal P_{\rm max} = 8.25\, \pm 1.0  \,\times  \,  10^{15} $ 
[J/s]. This is about $ 2.3 \pm 0.3$  times larger than $\mathcal
P_{\rm max,\odot}$. 

If mode masses are evaluated at the photosphere ($h=0$, $T=T_{\rm eff}$), we obtain
$\mathcal P_{\rm max} = 15.9  \, \pm 8.0 \, \times \,  10^{15} $
[J/s]. 
In that case this is about $\sim 4.4 \pm 2$  times larger than
$\mathcal P_{\rm max,\odot}$.

%For {\acenA} our modelling results in  $\mathcal P_{\rm max} = 8.82  \,\times
%\,  10^{15}$ [J/s],  this is about  2 times larger than $\mathcal P_{\rm max,\odot}$.

%\citet{Kevin06b} has inferred from the helioseismic
%constraints by \citet{Baudin05}  $\mathcal P_{\rm max,\odot} \simeq 
%4.5 \, \pm \, 0.4 \,\times \,  10^{15}$ [J/s] using mode mass obtained
%with an 1D model computed on the basis of 
%\citet{Gough77}'s non-local mixing-length theory of convection.
%If we use solar mode masses obtained with a patched model
%computed as described here, 

% and the excitation rates inferred from the observations
% give $\mathcal P_{\rm max} = 11.0  \,\times \, \pm 5.5 \,  10^{15} $
%[J/s].

As seen in Fig.~\ref{figH}, the frequency  where $\mathcal P$
 peaks  is $\sim$~2.4~mHz for {\acenA}. For comparison, in the solar case, $\mathcal P$ peaks  around
3.8~mHz. Clearly, the modes in {\acenA}  are excited at lower frequency compared
to the solar modes.
 Note also that the frequency domain where the derivation of $\mathcal P$
  is possible from the available seismic data is much smaller   for  {\acenA} than for the Sun. 
  This is obviously because the quality of the seismic data is much
  lower for {\acenA}  than for the Sun. 

\fig{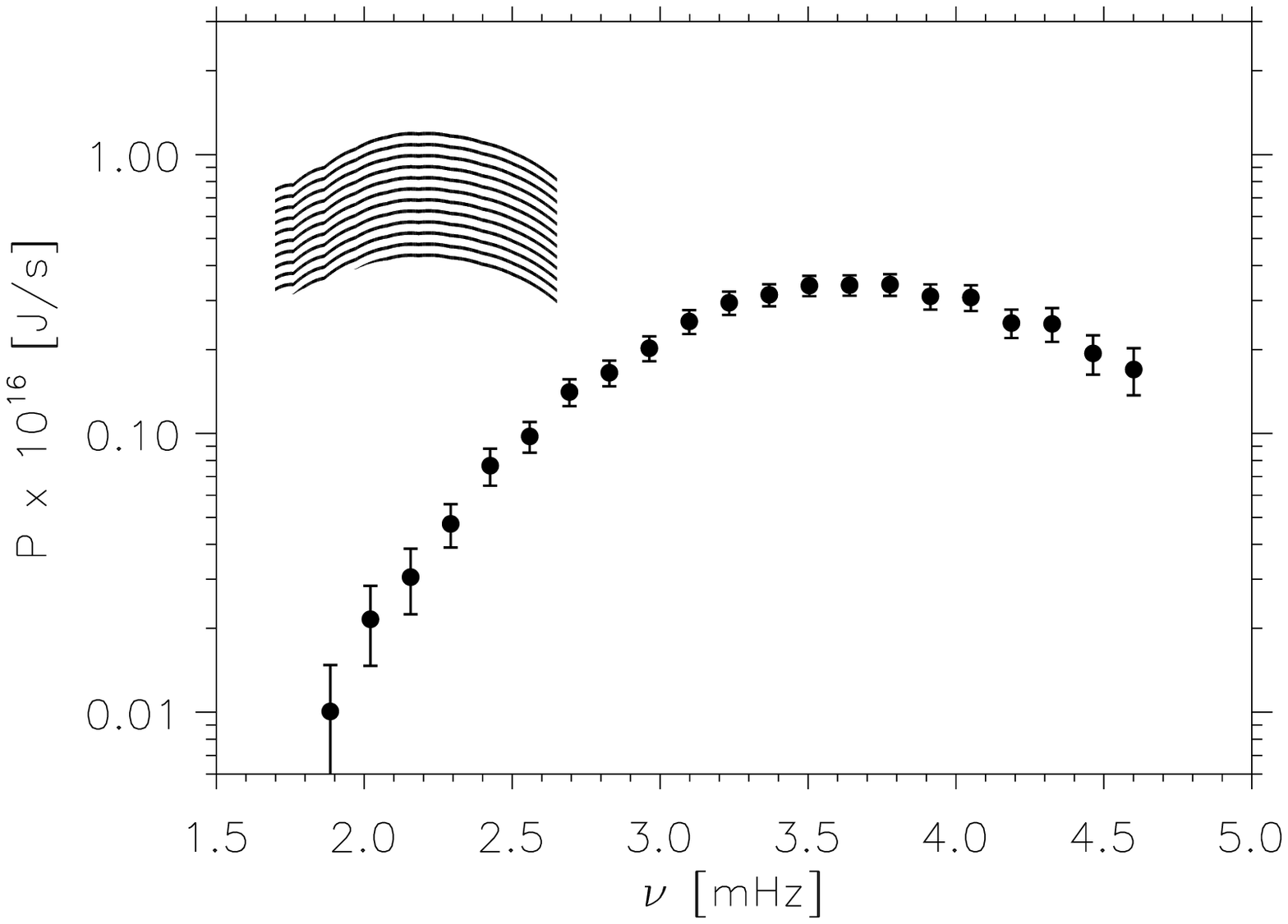}{
  Excitation rates ${\mathcal P}$ inferred from seismic data according to Eq.~\ref{Pobs}. 
  Filled circles correspond to the helioseismic constraints
  obtained by \citet{Baudin05}. The dashed area 
  represents the observed domain for the excitation rates derived
  for {\acenA}. 
 }{figH}

%---------------------------
\subsection{Excitation rates as a function of depth}

\begin{figure}[ht]
        \begin{center}
        \resizebox{\hsize}{!}{\includegraphics  {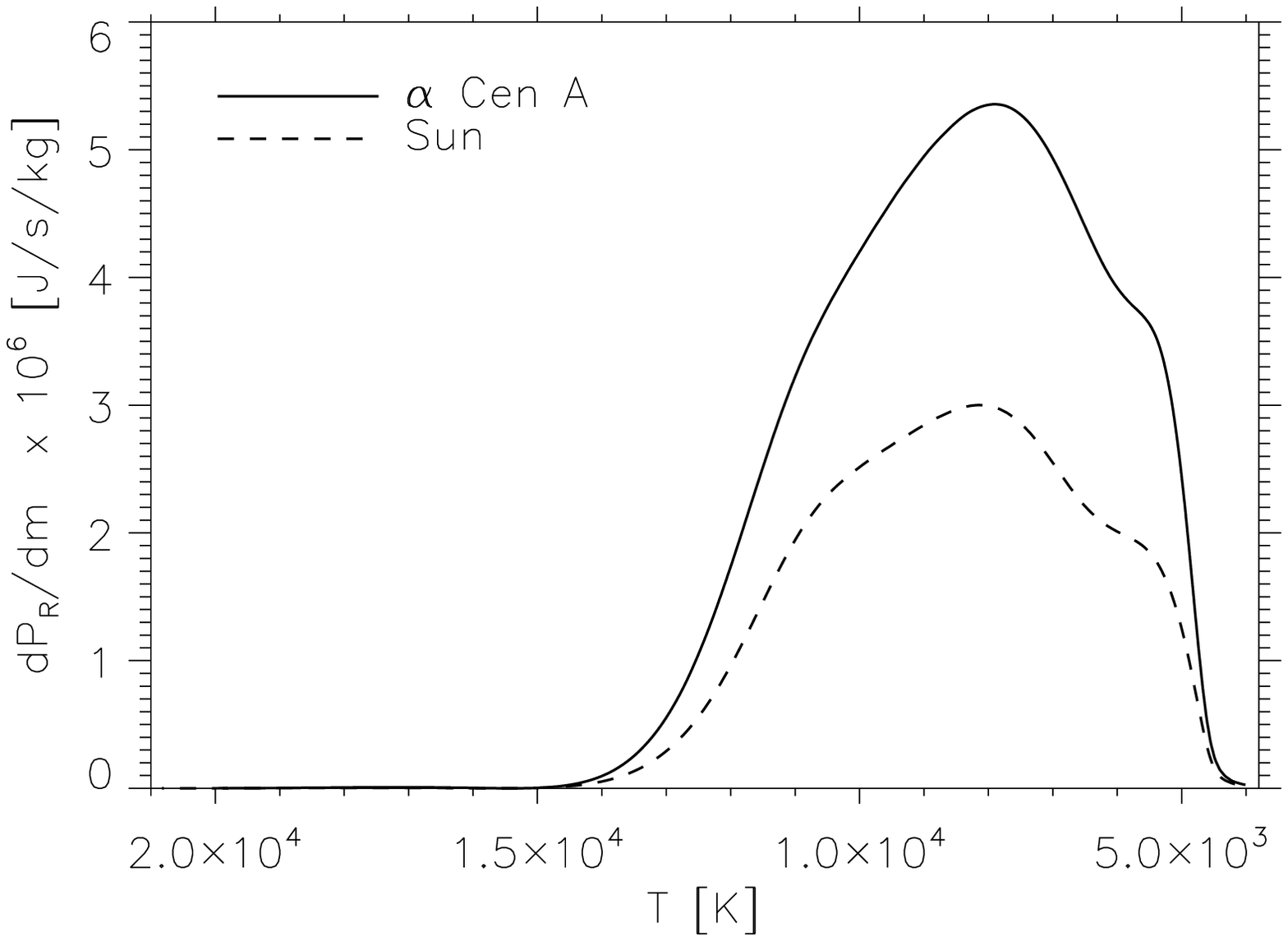}}
        \resizebox{\hsize}{!}{\includegraphics  {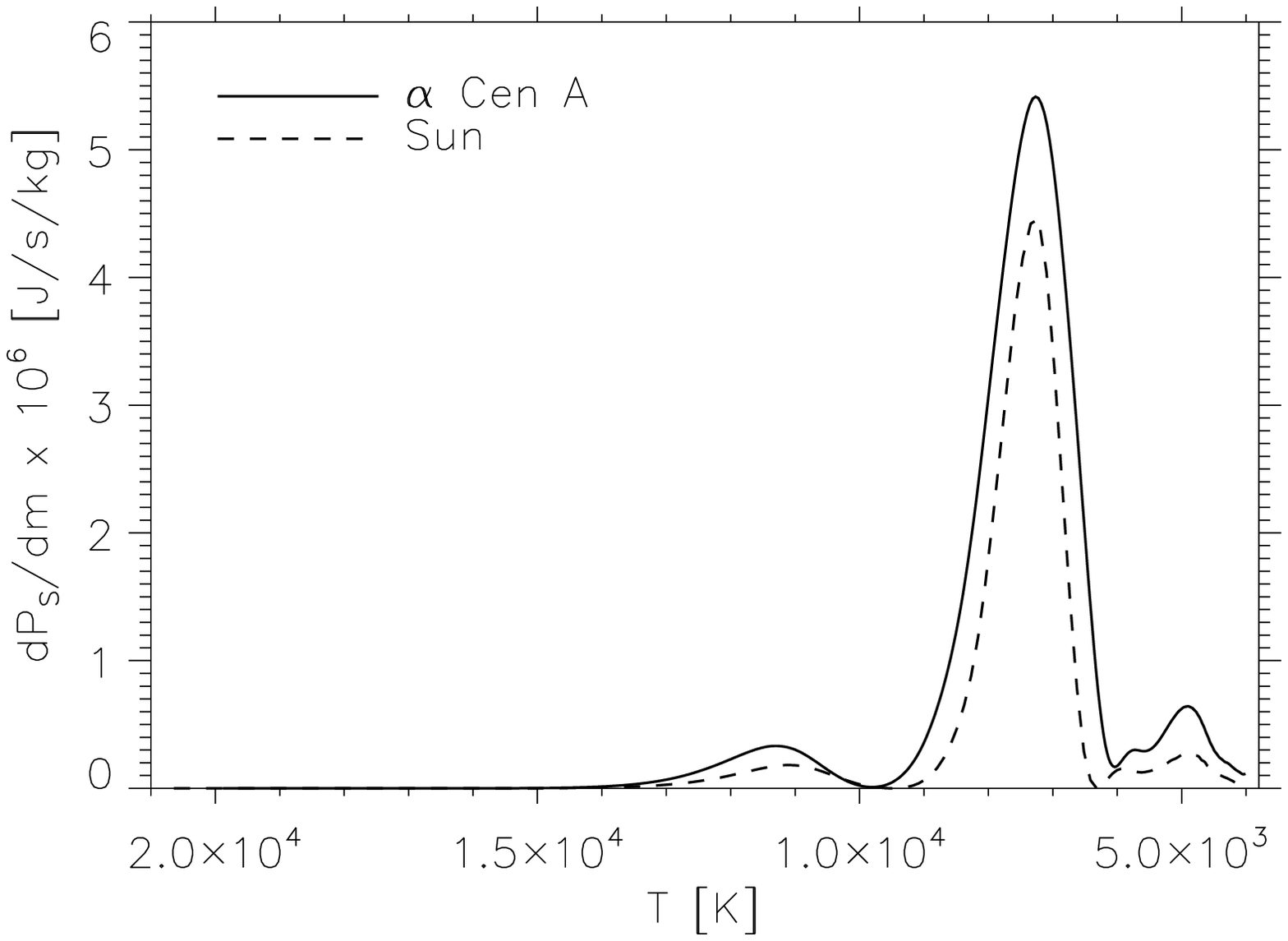}}
        \end{center}    
\caption{ {\bf Top:} The integrand $d{\mathcal P}_{\rm R}/dm$ (Eq.~\ref{power}) associated with the contribution
  of the Reynolds stress to the excitation is plotted as a function of
  the horizontally and temporally averaged temperature in the simulation box for the mode for which
  $\mathcal P$ is maximum. The solid line corresponds to
  the 3D simulation associated with 
  {\acenA} and the dashed line to the one associated with the Sun. 
{\bf Bottom:} as the top panel for $d{\mathcal P}_{\rm S}/dm$, the integrand  associated
with the contribution due to the entropy fluctuations. 
}
\label{figF}
\end{figure}

Fig.~\ref{figF} shows the  integrand $d{\mathcal P}/dm$ of the excitation
rates (Eq.~\ref{power},\ref{C2R_ref},\ref{C_S_ref})
as a function of the  temperature for the mode for which
  $\mathcal P$ is maximum in the Sun and in {\acenA}. The top panel shows the contribution
of the Reynolds stress ($d{\mathcal P}_{\rm R}/dm$) and the bottom panel the contribution of the
entropy fluctuations ($d{\mathcal P}_{\rm S}/dm$).
Excitation due to the Reynolds stress is maximum where the rms value of
velocity, $u$, peaks.
Excitation due to the entropy fluctuations is maximum
where $\tilde s$, the rms
 value of the entropy fluctuations, peaks. 
Fig.~\ref{figF} shows that the excitation rate is larger for {\acenA}
than for the Sun and occurs
over a slightly more extended region in {\acenA} than in the Sun.

The excitation due to the entropy fluctuations occurs in a more shallow
region compared to the Reynolds stress.
For {\acenA}, the relative contribution of the entropy fluctuations to
the total excitation is  $\simeq $~18\,\%, which is similar to the  Sun
($\simeq$~15~\%).
 Hence, in both cases the excitation due to the
Reynolds stress remains the dominant contribution.

%-------------------------------------------
\subsection{Differences in the characteristic properties of convection}

To summarize, we find that ${\mathcal P}$ is  significantly larger in
{\acenA} than in the Sun. 
Furthermore,  ${\mathcal P}$ peaks at lower frequency.
As shown below, all these seismic differences can be attributed to differences
in the characteristic properties of convection between {\acenA} and the Sun.

%-------------------------------------------
\subsubsection{Why $\mathcal P$ is larger for {\acenA} ?}
\label{Why_larger_P}

At a given layer, the power supplied to the modes by the \emph{Reynolds
stress} is  proportional  --~ per unit mass ~-- to $\rho_0
\, u^3 \,  \Lambda^4$  where $\Lambda$ is the characteristic eddy size and  $u$ the rms
 value  of the velocity \citep[see][]{Samadi00I}. The flux of the kinetic energy,
 $F_{\rm kin}$,  is  proportional to $\rho_0 \, u^3 $. 
Hence, the larger $F_{\rm kin}$  or $\Lambda$, the larger the driving by the Reynolds stress.

The power  supplied to the modes  
 by the so-called \emph{entropy source term} is proportional  --~ per unit mass 
~-- to  $\rho_0 \, 
u^3 \, \Lambda^4 \, \mathcal R^2 \,  / (\tau_\Lambda \, \omega_0)^2 $ where
$\omega_0$ is the mode frequency,   $\tau_\Lambda \sim \Lambda / u$
is the characteristic eddy turn over time, finally $\mathcal R
 \propto  F_{\rm conv} /
F_{\rm kin}$ where $F_{\rm conv} \propto w \, \alpha_s \,  \tilde s $
is the convective flux and $\tilde s$ is the rms of entropy fluctuations
\citep[see][]{Samadi05b}.
Note that the higher $\mathcal R$, the higher the \emph{relative}
contribution of the entropy source to the excitation.
Note also that the driving is maximum for mode frequency \citep[see, {\it e.g.},][]{Samadi00I}
\eqn{
\omega_0 \sim
2 \pi/ \tau_\Lambda \; .
\label{w0}
} 
Hence, at the mode frequency $\omega_0 \sim
2 \pi/ \tau_\Lambda$, the larger the ratio $ F_{\rm conv} /
F_{\rm kin} $  the larger the relative
contribution of the entropy source term to the total excitation rate.

%The Fig.~\ref{figD}  shows $u_z$ and $\tilde s$ as a function of temperature  for the 3D simulation
%associated with  {\acenA} and for the 3D  solar simulation considered by \citet{Kevin06a}.
%The maximum in  $u_z$ (resp. $\tilde s$) is $\sim$~10 \% (resp.  $\sim$~25
%\%) larger in the 3D simulation associated with {\acenA}  than in the
%solar one. 

As a summary, for both Reynolds stress  and entropy contributions, the larger the characteristic scale length
($\Lambda$) or the larger the kinetic energy ($F_{\rm kin}$), the
larger the excitation. Furthermore, the larger $\mathcal R$, the
larger the relative contribution of the entropy source term to the
excitation.
We study below the differences in $\Lambda$, $F_{\rm kin}$ and
$\mathcal R$ between the Sun and {\acenA}.

{\it Kinetic energy flux ($F_{\rm kin}$):}\\
The maximum in $u$ is up to $\sim$~10 \% larger in the 3D simulation associated with {\acenA}  than in the
solar one.  
However, the differences in the flux of kinetic energy, $F_{\rm kin}$,
between the 3D simulations associated with {\acenA} 
and the solar one are small ($\lesssim$~10 \%).
 This small effect on
    $F_{\rm kin}$ despite its cubic dependence on $u$ is
    due the lower $\rho_0$ for a layer with the same average
    $T$ in the simulation for {\acenA} as compared
    to the simulation for the Sun. The lower $\rho_0$ in
    turn is a consequence of the lower surface gravity
    of $\alpha$~Cen~A compared to the Sun. 

{\it  Relative contribution of the entropy source term:}\\
We also find  that  $\tilde s$ is  $\sim$~25 \% larger in the 3D simulation
associated with {\acenA}. However, the convective flux, $ F_{\rm conv}
\propto w \, \alpha_s \,  \tilde s $, in the 3D simulation associated
with {\acenA} is very close to that of the solar simulation. This
is not surprising since the two stars have almost the same effective
temperature.  Furthermore, as pointed out above, the differences in
$F_{\rm kin}$  between {\acenA} and the Sun are small.
As a consequence $\mathcal R   \propto  F_{\rm conv} /
F_{\rm kin} $ does not differ between {\acenA} and the Sun.
This explains   why the contribution of the entropy term relative to
the Reynolds stress  is  similar between {\acenA} and the Sun.

%\begin{figure}[ht]
%        \begin{center}
%        \resizebox{\hsize}{!}{\includegraphics  {plot/cmp_sun.0.eps}}
%        \resizebox{\hsize}{!}{\includegraphics  {plot/cmp_sun.1.eps}}
%        \end{center}    
%\caption{{\bf Top:} Root mean square of the vertical component
%  of the velocity as a function temperature. 
%  The solid line corresponds to the 3D simulation associated with
%  {\acenA} and the dashed line to the one associated with the Sun. 
%{\bf Bottom:} as the top panel for, $\tilde s$, the root mean square of the entropy
%fluctuations.
%}
%\label{figD}
%\end{figure}

{\it Characteristic scale length  ($\Lambda$) :}\\
Fig.~\ref{figE} shows the kinetic energy spectrum $E$ as a
function of the horizontal wavenumber $k$  and the scale length $\Lambda_k =
  2\,\pi/k$ for the layer where $u$ is maximum. 
As seen in Fig.~\ref{figE}, for the 3D simulation associated with {\acenA},
$E$ is maximum at a larger scale length compared to the solar
simulation.  Then, the eddies have a larger characteristic scale
length in {\acenA} than in the Sun.
This explains why  
the  excitation of p-modes is significantly stronger 
 for {\acenA} than for the Sun. 

Note that since the number of grid points is
     the same for both simulations, the {\acenA} simulation has
     a larger physical grid size, thus a smaller maximum wavenumber,
     hence the cut-off in the spectrum  occurs at a lower $k$. This
      explains the earlier drop-off of $E(k)$ for {\acenA}
     in Fig.~\ref{figE}. For that reason the high wavenumber part (beyond
     a $k$ value of about 15~Mm$^{-1}$, or a $\Lambda_k$ less than 0.4~Mm) should
     not be compared directly. 
     On the other hand, the scaling chosen in Fig.~\ref{figD} allows a
     direct comparison.

\fig{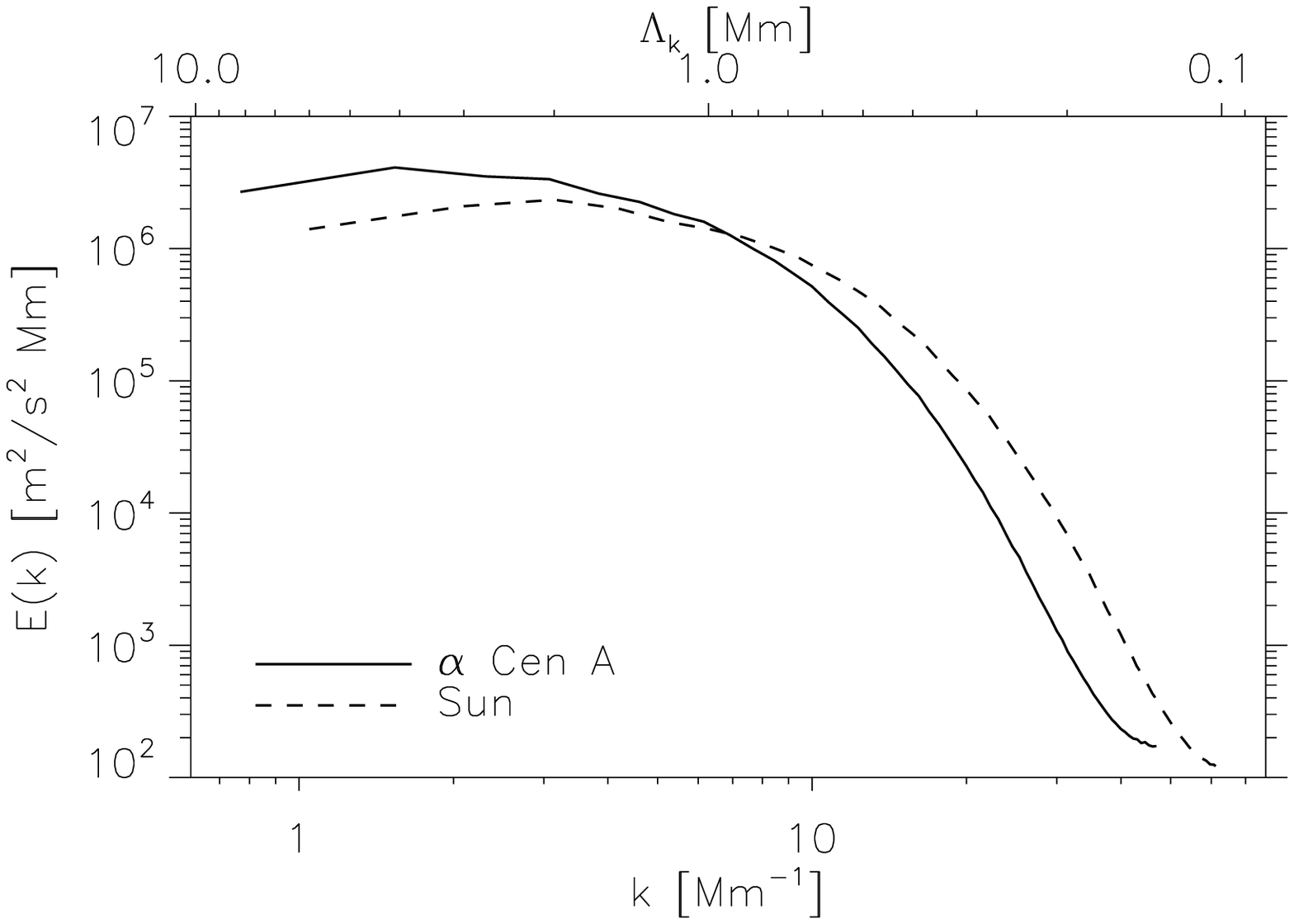}{
 The kinetic energy spectrum, $E$, as a function of the
  horizontal wavenumber $k$ (lower axis) and the scale length $\Lambda_k =
  2\,\pi/k$ (upper axis) for the layer where $u$ is
  maximum. The solid line corresponds to the 3D simulation associated with
  {\acenA} and the dashed line to the one associated with the Sun. 
}{figE}

We point out that the characteristic scale length, $\Lambda$, scales
as the pressure scale height. Indeed,
we have plotted in Fig.~\ref{figD} the kinetic energy spectrum, $E$, as a function of $k H_p$ where
$H_p$ is the pressure height at the layer where  $u$ is maximum.  
Except at small scale lengths, we see that the $k$-dependency of the
spectrum  is almost the same between
the simulation associated with {\acenA} and the solar one.

The ratio between $H_p^{\rm \alpha\, cen\, A}$ and  $H_p^\odot$ (
$H_p^{\rm \alpha\, cen\, A}/ H_p^\odot \simeq$
1.38) is very close to the ratio $g_\odot /g_ {\rm \alpha\,cen\, A}$ ( $\simeq$  1.36).
This is obviously related to the fact that $H_p = P/\rho g \propto
T/g$.
Accordingly, since ${\cal P} \propto \Lambda^4$ (see above), 
we then have ${\cal P}/{\cal P}_\odot \propto (\Lambda/\Lambda_\odot)^4
\propto ( H_p^{\rm \alpha\, cen\, A}/ H_p^\odot)^4 \propto ( g_{\rm \alpha\, cen\, A}/ g_\odot)^4~\sim 3.4$. For comparison
 excitation rates computed for {\acenA} are two times larger than in the Sun.

\fig{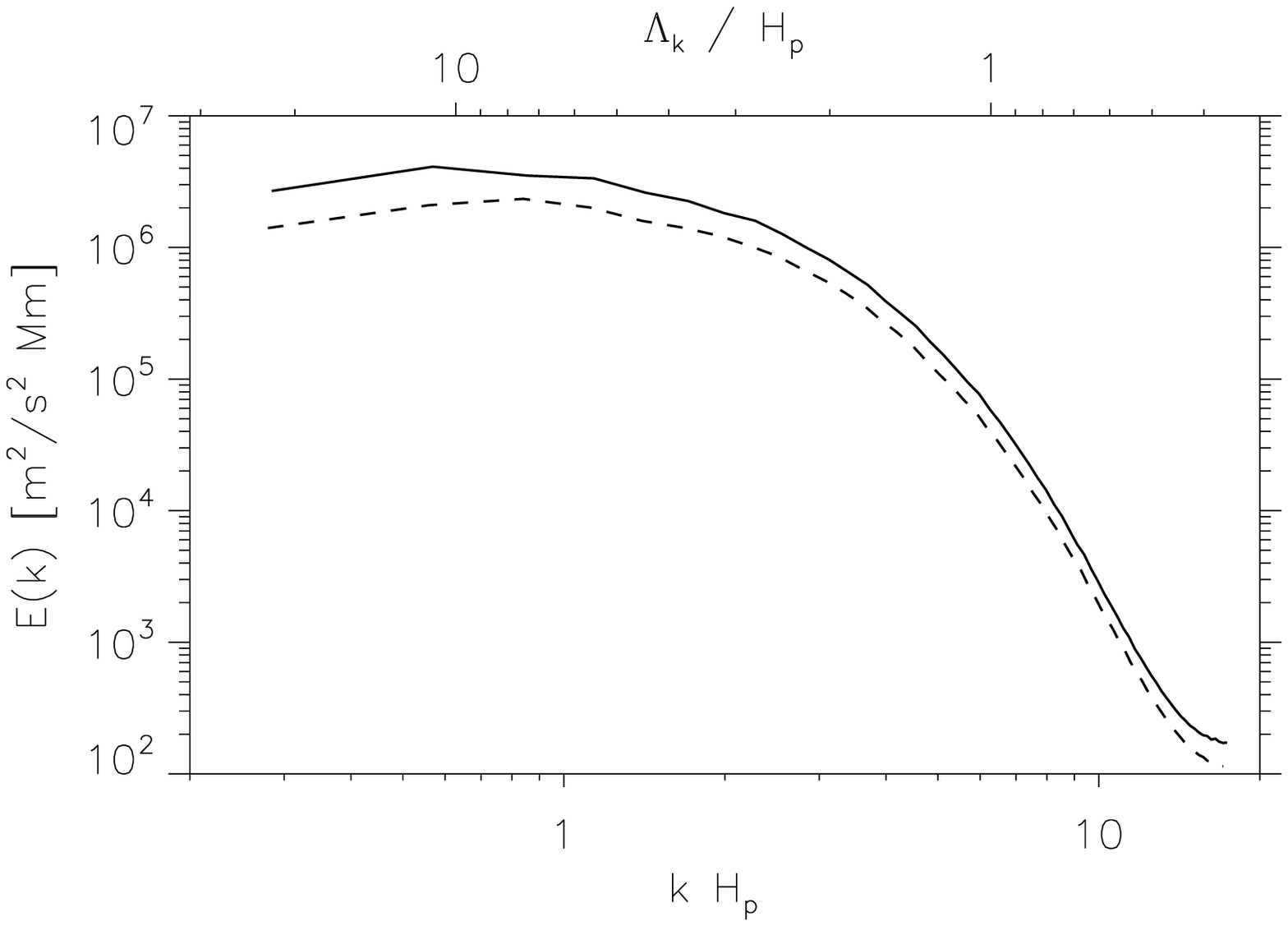}{
 The kinetic energy spectrum, $E$, as a function of 
   $k\, H_p$ (lower axis) and  $\Lambda_k / H_p$ (upper axis) for the layer where $u$ is
  maximum. The lines used have the same meaning as in
  Fig.~\ref{figE}. 
}{figD}

%For the 3D simulation associated with {\acenA},
%$E$ is maximum at the scale length $\Lambda_k \sim$~4~Mm while, for
%the solar 3D simulation, it is maximum at the length $\Lambda_k
%\sim\,$2~

%-------------------------------------------
\subsubsection{Why $\mathcal P$ peaks at lower frequency ?}

The characteristic  eddy turn-over time can be estimated as the  quantity
$\tilde {\tau} \sim L_h /u $ where $L_h$ is the horizontal extend  of the 3D
model and $u$ the velocity at a given layer. 
At the layer where $u$ is maximum, we find that $\tilde {\tau}$,  evaluated at the
layer where $u$ peaks, is larger
in {\acenA} ($\sim$ 30 minutes) than in the Sun ($\sim$ 23 minutes). 
This explains the fact that  for {\acenA} $\mathcal P$
peaks   at lower frequency than  in the Sun ($\omega_0 \sim 2\pi /
\tilde {\tau}$, cf. Eq.~\ref{w0}).  

Note that  both   $u$  and  $\Lambda$  are larger for {\acenA}  than
for the Sun. 
However, the net result is a larger  $\tilde {\tau} $ for {\acenA}.

%-------------------------------------------
\subsubsection{Interpretation}

The differences in characteristics of convection between {\acenA}
and the Sun can be understood as follows:
 as seen in Sect~\ref{Why_larger_P}, the characteristic size
  $\Lambda$ is mainly controlled by
  $H_{\rm p} \propto T/g$ (for a given composition). 
The surface gravity for {\acenA} is  $\sim$~35\,\% times smaller compared to the Sun
 while the effective temperature is very similar to that of the Sun. 
Consequently, $\Lambda$  is larger than in the
Sun. 
Furthermore because of the lower gravity, 
the density at the photosphere is smaller than in the
Sun.
Hence, in order to transport by
convection the same amount of energy per unit surface area, the
convective cells must have larger speed ($u$).

%Now,  because of the lower density at the surface, the granules are
%optically less thick in {\acenA}. Hence, their radiative losses are larger as
%shown by the larger values of $\tilde s$.
%However the increase of $u$ and $\tilde s$ compensate the decrease
%in the density such that $F_{\rm kin}$ and $F_{\rm conv} \propto
%w \alpha_s  \tilde s$ are similar between {\acenA} and the Sun.
%This is why the relative contribution of the entropy source term is
%similar between the Sun and {\acenA} (we recall that  $\mathcal R \propto F_{\rm conv} /
%F_{\rm kin}$).

%%%%%%%%%%%%%%%%%%%%%%%%%%%%%%%%%%%%%%%%%%
\section{Discussion}
\label{discussion}

%-------------------------------------
\subsection{Effect of chemical composition}

The star {\acenA} has an iron to hydrogen abundance of [Fe/H]=0.2 \citep[see][]{Miglio05}.
The 3D simulation considered here has a solar abundance.
Preliminary work tends to show that, at the given effective
temperature, a 3D simulation with a metal abundance 10
times smaller than the solar one results in  mode excitation rates
$\sim 2$ times smaller. 
This can be understood as follows: 
%at low metallicity, the mean opacity is smaller than at solar
%metallicity. In the superadiabatic region, where 
%convection is inefficient, 
the radiative flux is larger for a low metallicity 
than in a medium with a solar metallicity. 
In that case, to transport the same amount of energy,  
convection is less  vigourous ({\it i.e.} lower flux of kinetic energy, $F_{\rm  kin}$), 
%need not be as much vigorous, therefore 
leading to a lower efficiency of the driving.
If we extrapolate this preliminary result, we can expect  that mode
excitation rates ought to increase with [Fe/H]. A quantitative estimate
of the expected increase must however be performed, in particular for {\acenA},  which will 
 require to compute a 3D simulation with a non-solar abundance
representative of the surface layers of the star (in progress).

%-------------------------------------
\subsection{Estimation of mode mass}

%By default the mode masses have been evaluated at the photosphere. 
Mode masses must be evaluated at the layer in the atmosphere
where the acoustic modes are predominantly measured. 
The result of a comparison with seismic
  constraints significantly depends  on the effective heights $h$ where mode 
  masses are evaluated.  
Indeed, we plot in Fig.~\ref{figG}  the ratio $\mathcal P / \mathcal M$  for mode
masses evaluated at different heights $h$ in the atmosphere, namely
from $h=0$ (the photosphere) up to the top of the simulated domain ($h
\simeq 1000$~km). This ratio is compared to the quantity $  2 \pi\,
\Gamma  \,  (v/S_0)^2 $ obtained
from  the seismic constraints (Eq.~\ref{Pobs}).  
For $h \gtrsim$~600~km ({\it i.e.} for optical depth $\lesssim$~0.005), the ratio $\mathcal P / \mathcal M$ is outside
the observational domain. 

\fig{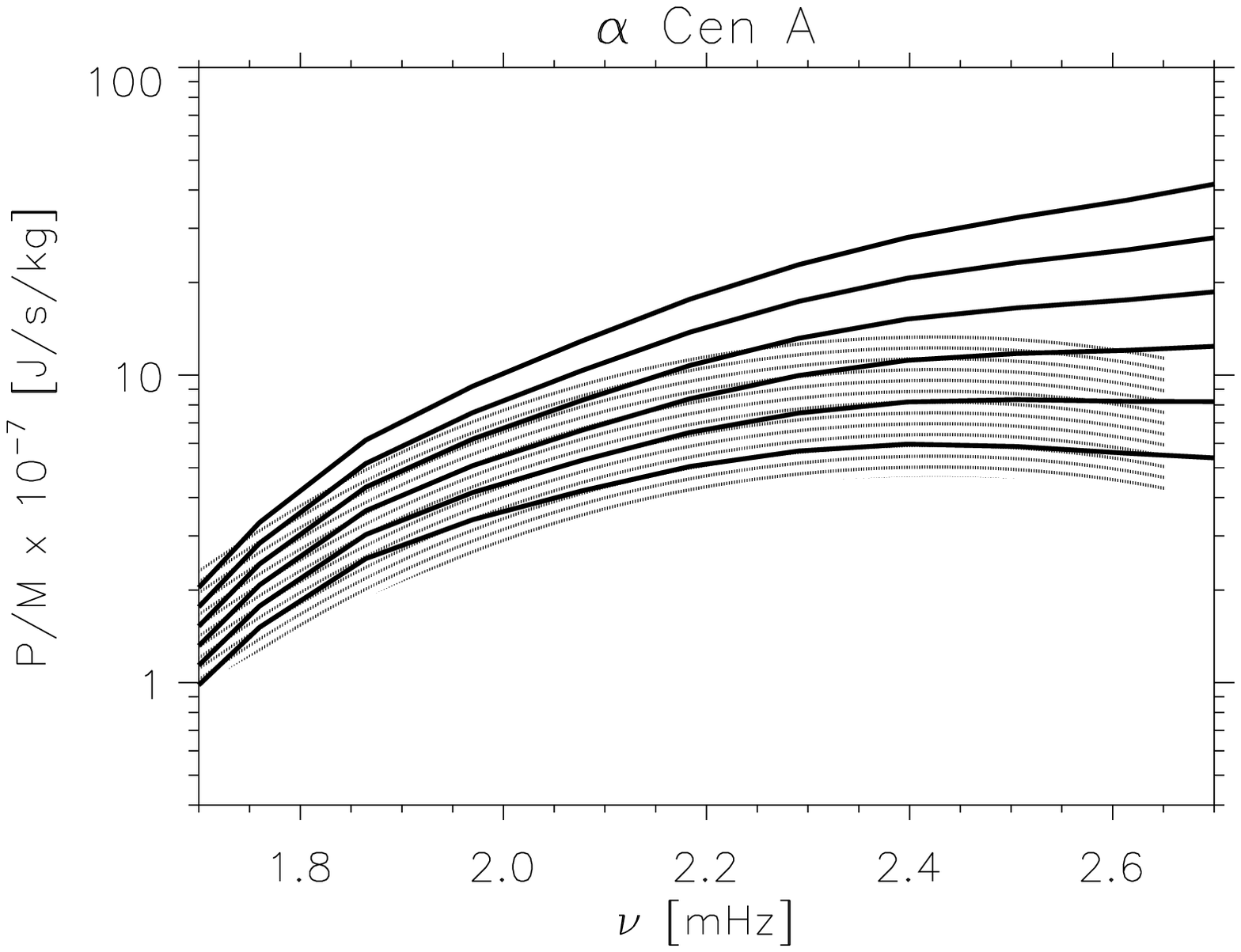}{
 Ratio of the rates ${\mathcal P}$ at which energy is injected into
  the $p$-modes   to the mode masses (${\mathcal
    M}$) for {\acenA}. The dashed area represents the observed domain for ${\mathcal
  P}/{\mathcal M} = 2 \pi \, \Gamma  (v/S_0)^2 $  as a function of
  $\nu$ (see Sect.~\ref{constraints}). The solid  lines correspond to the ratio  ${\mathcal P}/
  {\mathcal M}$ where the excitation rates, $\mathcal P$, are calculated according to
  Eq.~\ref{power} and the mode masses,  $\mathcal M$, are evaluated at different heights $h$ above
the photosphere. The lower curve corresponds to the photosphere
($h=0$) and the upper curve to the top of the atmosphere ($h=$~1000~km). The step
in $h$ is 200~km. 
}{figG}

%Helioseismic constraints obtained from spectrometry are usually obtained using
%spectrometric measurements at a given spectral line.
% However as pointed out by \citet{Baudin05}, even in the case of
%  helioseismic measurements, it is not trivial to evaluate the  height
%  $h$ above the photosphere where the observations 
%are predominantly performed.  

Seismic observations of {\acenA} were performed using UCLES and UVES
spectrographs. UVES and UCLES use a similar technique to measure the
acoustic modes (T. Bedding, private communication). 
As other spectrographs dedicated to stellar seismic measurements, 
the UCLES instrument uses  a large number of spectral lines in order to reach a high 
enough signal to noise ratio. 
In the case of stellar seismic measurements it is then  more difficult than
for helioseismic observations to estimate the effective height $h$ \citep[for
the solar case see, {\it e.g.},][]{Baudin05}.
A recent work by \citet{Kjeldsen08} allows us to estimate the value
for an effective $h$. 
Indeed, the authors have found that   
 solar modes measured with the UCLES
spectrograph have amplitudes slightly smaller than those measured by
the BiSON network.  
The instruments of the BiSON network use the potassium (K)
spectral line which is formed at an optical depth $\tau_{\rm 500~nm}
\simeq 0.013$
\citep[see][]{Houdek06}. 
 \citet{Kjeldsen08}'s results then suggest that acoustic modes measured by UCLES are 
measured at an effective height ($h$) slightly below the formation depth of the K line,  {\it i.e.}
at optical depth slightly above  $\tau_{\rm 500~nm} \simeq 0.013$. 
 Accordingly, we have here evaluated  
  the mode masses at that optical depth, which corresponds to $h$=470~km.

%Therefore, in contrast with HARPS instrument, we can not
%known whether UCLES measurements  are performed rather above or below the
%formation depth of the potassium line.

A more rigorous approach would be  to compute an effective mode mass by
  weighting appropriately the different mode masses associated with the different
  spectral lines that contribute to the seismic measure.
In order to infer accurate mode excitation rates from
the seismic data of {\acenA}, the mode masses
representative for the observation technique  \emph{and} the 
  spectral lines of {\acenA} must be derived.
This is out of scope here.

%%%%%%%%%%%%%%%%%%%%%%%%%%%%%%%%%%%%%%%%%%
%%%%%%%%%%%%%%%%%%%%%%%%%%%%%%%%%%%%%%%%%%
\section{Conclusions}
\label{conclusion}

Theoretical estimations for  the   energy supplied per unit of time by turbulent convection ($\mathcal P$)  to {\acenA} acoustic modes
 have been compared to  values obtained from observations. 
 This allows us to draw the following conclusions:

%-------------------------------------
\subsection{Differences with the Sun}
 
Although  {\acenA} has an effective temperature very close to that
of the Sun, 
we find here that  the p-mode excitation rates $\mathcal P$ inferred
  from the seismic constraints obtained for {\acenA} 
  are  about two times larger than in the Sun.
These differences are  attributed to the fact that the eddies in {\acenA} have a
larger characteristic size ($\Lambda$) than in the Sun. This is
related to the fact that {\acenA} has a smaller surface gravity.

Furthermore, the p-mode excitation rates for {\acenA} are maximum at lower frequencies than
in the Sun. This  behaviour is related to the fact that the eddies have
a longer turn-over time as a result of a larger $\Lambda$.

The seismic characteristics of the p-modes detected in {\acenA} 
 significantly differ from that of the Sun. They can therefore provide additional
constraints on the model of stochastic excitation.

%-------------------------------------
\subsection{Inferred versus modelled  excitation rates}

 Our modelling gives rise to  
  excitation rates within the error bars associated with the observational
  constraints. We stress that this modelling was undertaken for {\acenA} {\it independently}
  from the solar case, {\it i.e.} without  using a formulation
  fitted on the helioseismic data  as it is the case, for instance,
  in the case of the Sun in \citet{Chaplin05} or in the case of {\acenA}
  in \citet{Houdek02b}.  
The seismic constraints from {\acenA} then provide a clear validation of
the basic underlying physical assumptions included in the theoretical
model of stochastic excitation, at least for stars not too different
from the Sun.

%-------------------------------------
\subsection{Constraints on the description of turbulence: eddy-time
  correlation}

We find that our theoretical estimations of $\mathcal P$,  which assume 
 a Lorentzian eddy-time correlation function ($\chi_k$) and the 
Closure Model with Plumes (CMP) proposed by \citet{Kevin06a}, lie  in
the observed domain.
%between the
%two different sets of seismic constraints. 
On the other hand, when a Gaussian function  is chosen for $\chi_k$,
$\mathcal P$ is significantly underestimated.
 The comparison with the seismic data for {\acenA} confirms 
  the results  for  the solar case obtained by  \citet{Samadi02II} 
  that  $\chi_k$ significantly departs  from a Gaussian. 
As in  \citet{Samadi02II}, we attribute the departure of $\chi_k$
from a Gaussian to diving plumes ({\it i.e.} 
down-flows), which are more turbulent than granules ({\it i.e.} the
up-flows).
 This result confirms  that a
  Lorentzian function is a more adequate description for the eddy-time
  correlation  than a Gaussian.

%This departure from the Gaussian $\chi_k$  has the consequence to spread
%more energy above the characteristic frequency of the most energy
%bearing eddies  \citep[see][]{Samadi02II}. 

%-------------------------------------
\subsection{Constraints on the modelling of turbulent convection in the
  equilibrium stellar model}

Calculations involving eigenfunctions computed on the
basis of a global 1D model that includes a realistic description of
the outer layers of the star (taken from 3D
simulations) reproduce  much better (see Fig.~\ref{figA}) 
the seismic data than calculations that use eigenfunctions computed with 
a  standard stellar model built with  the 
mixing-length theory (MLT) and ignoring turbulent pressure. 
This is because a model that includes turbulent pressure results in
larger mode masses ${\cal M}$ than a model which ignores turbulent
pressure. 
This can be understood as follows. Within the super-adiabatic region, 
a model that includes turbulent pressure provides an additional
support against gravity and hence has a lower gas pressure
and density (see Fig.~\ref{figC}) than a model that does not include
turbulent pressure.
As a consequence,  mode inertia  (and hence mode masses) are then larger in a
 model that includes turbulent pressure. 

These conclusion are similar to that obtained in the Sun. 
Indeed, the mode masses considered by \citet{Kevin06b} in the case
of the Sun were obtained with a 1D model computed using \citet{Gough77}'s  non-local
 mixing-length formulation of convection. The model thus includes
 turbulent pressure.
We do not observe significant differences between excitation rates
obtained  with this  non-local model and those obtained with  a
``patched'' solar computed as described here in the case of {\acenA}.
 On the other hand, excitation
 rates computed with mode masses obtained with a ``standard''  solar model
 (that is with no turbulent pressure included) or with a model in which turbulent
 pressure is included according to the mixing-length theory 
%(see  eg. XX) 
under-estimate significantly the helioseismic constraints.

% At $\nu_{\rm max} \simeq $~3.8~mHz, where ${\cal P}$ is maximum,
% the mode masses  used by \citet{Kevin06b} are 
% $\sim$~25~\% larger compared  to thoses obtained with a  patched solar
% model.  

 These results tell that one must compute mode
  masses from 1D models that include turbulent pressure 
 using a 3D hydrodynamical model or using a
  non-local description of convection.

%-------------------------------------
\subsection{Need for improved data sets}

As shown by \citet{Samadi02II} in the case of the Sun, contribution of
the entropy fluctuations to the excitation cannot be neglected.
Furthermore, recently, \citet{Kevin06b} have shown that theoretical
calculations based on the CMP result in a better agreement with the
helioseismic constraints than those based on the Quasi-Normal
Approximation (QNA).

However, in the case of {\acenA}, differences between theoretical
calculations that use the CMP  and those based on the
QNA (see Fig.~\ref{figB}) as well as differences between
calculations including  driving by  entropy fluctuations and
those that do not include it (not shown), are of the same order as the
observational uncertainties associated with the two data sets. 
%Note that the difference between the two different sets of seismic
%constraints are found larger than the differences between the 
%theoretical calculations in question. 
%This is a consequence of the large difference between the two estimates for the mode
%line widths $\Gamma$. %, which discrepancy remains to be understood. 
The present seismic constraints  therefore are  unable to
discriminate between these assumptions.  This emphasizes the
need for more accurate seismic data for {\acenA}.

%%%%%%%%%%%%%%%%%%%%%%%%%%%%%%%%%%%%%%%%%%
\begin{acknowledgements}
FK's work was possible thanks to a one month grant provided by Observatoire de Paris.
We thank T. Bedding for  providing us  the
amplitude spectrum of {\acenA} obtained from the UVES and UCLES
spectrographs.
We thank {\AA}. Nordlund, R.F. Stein and R. Trampedach for  making
their 3D simulation code available.
Finally, we thank the referee (M. Steffen) for his useful remarks.
\end{acknowledgements}

\bibliographystyle{aa}
%\bibliography{/home/reza/redac/biblio.bib}

\end{document}